\documentclass{revtex4}
\input epsf
\usepackage{amsmath,amssymb}
\usepackage{graphicx}

\draft

\begin{document}
\titlepage
\title{Gravitational potential in Palatini formulation of modified gravity}
\author{Xin-He Meng$^{1,2,3}$ \footnote{xhmeng@phys.nankai.edu.cn}
 \ \ Peng Wang$^1$ \footnote{pewang@eyou.com}
} \affiliation{1.  Department of Physics, Nankai University,
Tianjin 300071, P.R.China \\2. Institute of Theoretical Physics,
CAS, Beijing 100080, P.R.China \\3. Department of Physics,
University of Arizona, Tucson, AZ 85721}

\begin{abstract}
General Relativity has so far passed almost all the ground-based
and solar-system experiments. Any reasonable extended gravity
models should consistently reduce to it at least in the weak field
approximation. In this work we derive the gravitational potential
for the Palatini formulation of the modified gravity of the L(R)
type which admits a de Sitter vacuum solution. We conclude that
the Newtonian limit is always obtained in those class of models
and the deviations from General Relativity is very small for a
slowly moving source.
\end{abstract}

\maketitle

Recently, some attempts have been made to explain the observed
cosmic acceleration \cite{Perlmutter} by modifying the
Einstein-Hilbert action [2-17]. Those include the $1/R$ gravity
\cite{CDTT} which may be predicted by String/M theory
\cite{Odintsov-string}, the $1/R+R^2$ \cite{Odintsov-R2} and $\ln
R$ \cite{Odintsov-lnR} gravity which intend to explain both the
current acceleration and early time inflation. Generally, those
models have two \textbf{inequivalent} formulations: the metric
formulation (second order formulation) and the Palatini
formulation (first order formulation) \cite{Vollick, Flanagan,
Wang1, Wang2, Wang3}. There are many interesting features in the
Palatini formulation of those models, e.g. the absence of the
instability \cite{Wang1} appeared in the metric formulation
\cite{Dolgov} (However, this instability in metric formulation may
be resolved by a $R^2$ term \cite{Odintsov-R2} or conformal
anomaly induced terms \cite{Odintsov-mini}), the universality of
the field equations for vacuum \cite{Volovich} or source with
constant trace of the energy-momentum tensor \cite{Rubilar}. There
is also a third alternative to the above two formulations
\cite{Flanagan2, Wang4}. In those works, it has been shown that
the simplest proposal of $1/R$ gravity in metric formulation is in
conflict with solar system gravitational experiments \cite{Chiba}
and the Palatini formulation is in conflict with electron-electron
scattering experiments \cite{Flanagan}. But it is still worth
continuing to explore whether some phenomenological model of this
type or its variants can easily accommodate the data. Thus it is
suitable to explore some basic features of this type of models at
this time. Specifically, any reasonable gravity theory should
reduce to Newton gravity for slowly moving weak source. The
condition of having a Newtonian limit for the metric formulation
is explored in Ref.\cite{Dick}. In this paper, we will discuss the
weak field expansion of Palatini formulation. We will follow the
sign conventions of Ref.\cite{Flanagan}.

In general, when handled in Palatini formulation, one considers
the action to be a functional of the metric $\bar{g}_{\mu\nu}$ and
a connection $\hat{\bigtriangledown}_{\mu}$ which is independent
of the metric. The resulting modified gravity action can be
written as
\begin{equation}
S[\bar{g}_{\mu\nu},
\hat{\bigtriangledown}_{\mu}]=\frac{1}{2\bar\kappa^2}\int
d^4x\sqrt{-\bar{g}}L(\hat{R})+S_M,\label{1}
\end{equation}
where $\bar\kappa$ is a constant with dimension (mass)$^{-1}$ that
will be specified below, $\hat{R}_{\mu\nu}$ is the Ricci tensor of
the connection $\hat{\bigtriangledown}_{\mu}$,
$\hat{R}=\bar{g}^{\mu\nu}\hat{R}_{\mu\nu}$ and $S_M$ is the matter
action. For the theory (1) to explain the current cosmic
acceleration, the vacuum solution is necessarily de Sitter. Thus,
when studying its Newtonian limit, we must expand around de Sitter
background. The Newtonian limit of the metric formulation of this
sort of modified gravity was studied in Ref.\cite{Dick} and it
concluded that the existence of a weak field approximation around
the de Sitter background can be achieved by imposing the condition
$L''(R_0)=0$ where $R_0$ is the vacuum solution in metric
formulation. All the current existed model that may explain the
current cosmic acceleration, namely, the $1/R$ and $\ln R$ model,
do not satisfy this condition. And it is shown in
Ref.\cite{Flanagan} that the condition for the existence of an
equivalent scalar-tensor description of the theory (1) is just
$L''\neq0$. The absence of such an equivalent description is a
strong indication that the original theory is unphysical
\cite{Magnano}. Thus if the condition for the existence of
Newtonian limit in Palatini formulation is the same as the metric
formulation, the whole framework of explaining cosmic acceleration
in Palatini formulation is doubtable. Fortunately, as we will show
in this work. This is not the case. The theory (1) always has
Newtonian limit and the deviations from the Newtonian potential
with a standard Yukawa term (since we are working in de Sitter
background) are very small for a slowly moving source.

First, let us give the field equations of (1). See
Ref.\cite{Vollick} for details.

Varying the action with respect to $\bar{g}_{\mu\nu}$ gives
\begin{equation}
L'(\hat R)\hat R_{\mu\nu}-\frac{1}{2}L(\hat R)\bar
g_{\mu\nu}=\bar\kappa^2 T_{\mu\nu},\label{2}
\end{equation}
where $T_{\mu\nu}$ is the energy momentum tensor and is given by
\begin{equation}
T_{\mu\nu}=-\frac{2}{\sqrt{-\bar g}}\frac{\delta S_M}{\delta \bar
g^{\mu\nu}}.\label{2.3}
\end{equation}

Varying with respect to the connection, one can find that $\hat
R_{\mu\nu}$ is related to $\bar R_{\mu\nu}$ by
\begin{equation}
\hat R_{\mu\nu}=\bar R_{\mu\nu}+\frac{3}{2}(L')^{-2}\bar\nabla
_{\mu}L'\bar\nabla _{\nu}L'-(L')^{-1}\bar\nabla _{\mu}\bar\nabla
_{\nu}L'-\frac{1}{2}(L')^{-1}g_{\mu\nu}\bar\square L'\label{Ricci}
\end{equation}
and contracting with $\bar g_{\mu\nu}$ gives
\begin{equation}
\hat R=\bar R+\frac{3}{2}(L')^{-2}(\bar\bigtriangledown_\mu
L)^2-3(L')^{-1}\bar\square L'\label{scalar}
\end{equation}
where $\bar R_{\mu\nu}$ is the Ricci tensor with respect to $\bar
g_{\mu\nu}$ and $\bar R=\bar g^{\mu\nu}\bar R_{\mu\nu}$.

By contracting Eq.(\ref{2}) with $\bar g_{\mu\nu}$, we can solve
$\hat R$ as a function of $T$:
\begin{equation}
L'(\hat R)\hat R-2L(\hat R)=\bar\kappa^2 T \label{R(T)}
\end{equation}
where a prime denotes differentiation with respect to $\hat R$.
Thus (\ref{Ricci}), (\ref{scalar}) do define the Ricci tensor with
respect to $\hat{\bigtriangledown}_{\mu}$ and $L(\hat R)$ in
Eq.(\ref{2}) is actually a function of $T$. Specifically, we
denote the vacuum solution by $\hat R_0=R_0$ and thus $\hat
R^{(0)}_{\mu\nu}=\bar g_{\mu\nu}R_0/4$. By Eq.(\ref{Ricci}), $\bar
R^{(0)}_{\mu\nu}=\hat R^{(0)}_{\mu\nu}$ and thus $\bar R_0=\hat
R_0=R_0$.

The aim of this paper is to treat the approximation in which
gravity is "weak". In the context of modified gravity, this means
that the spacetime metric is nearly de Sitter. In application of
gravity theory, this is an excellent approximation except for
phenomena dealing with strong gravitational fields such as black
holes and large scale structure of the universe. Thus in presence
of some source, we divide the metric into two parts
\begin{equation}
\bar g_{\mu\nu}=\bar g^{(0)}_{\mu\nu}+h_{\mu\nu}\label{}
\end{equation}
where $\bar g^{(0)}_{\mu\nu}$ is the de Sitter vacuum solution of
the field equations and $h_{\mu\nu}$ represents deviations which
vanish at infinity. The first order variation of $\bar R_{\mu\nu}$
is given by the Palatini identity
\begin{equation}
\delta\bar R_{\mu\nu}=\frac{1}{2}(\bar\nabla_\mu\bar\nabla^\rho
h_{\rho\nu}+\bar\nabla_\nu\bar\nabla^\rho
h_{\rho\mu})+\frac{1}{3}R_0h_{\mu\nu}-\frac{1}{12}R_0\bar
g^{(0)}_{\mu\nu}h-\frac{1}{2}\bar\square
h_{\mu\nu}-\frac{1}{2}\bar\nabla_\mu\bar\nabla_\nu h\label{11}
\end{equation}
where $h=h_{\mu\nu}\bar g^{(0)\mu\nu}$.

Since in the vacuum, we have $\hat R^{(0)}_{\mu\nu}=\bar
R^{(0)}_{\mu\nu}$, the relation between perturbations of those two
quantities can be read off from Eq.(\ref{Ricci}):
\begin{equation}
\delta \hat R_{\mu\nu}=\delta\bar
R_{\mu\nu}+\frac{3}{2}(L')^{-2}\bar \bigtriangledown_\mu
L'\bar\bigtriangledown_\nu
L'-(L')^{-1}\bar\bigtriangledown_\mu\bar\bigtriangledown_\nu
L'-\frac{1}{2}(L')^{-1}\bar g^{(0)}_{\mu\nu}\bar\square
L'\label{12}
\end{equation}
and contracting with $\bar g^{(0)\mu\nu}$ gives
\begin{equation}
\delta\hat R=\delta\bar
R+\frac{3}{2}(L')^{-2}(\bar\bigtriangledown_\mu
L)^2-3(L')^{-1}\bar\square L'\label{13}
\end{equation}

A subtlety in expansion with respect to de Sitter background is
that $\hat R\neq R_0+\delta\hat R$, instead, by writing $\hat
R=(\bar g^{(0)\mu\nu}+h^{\mu\nu})(\hat R^{(0)}_{\mu\nu}+\delta
\hat R_{\mu\nu})$ explicitly, we can find that $\hat
R=R_0-\frac{R_0}{4}h+\delta\hat R_{\mu\nu}$, where
$h=h_{\mu\nu}\bar g^{(0)\mu\nu}$. Without the $h$ term, the
resulting equation will not be gauge-invariant. Specifically, this
term should be added to the expansion equation in Ref.\cite{Dick},
but its conclusion is unchanged.

The first order expansion of Eq.(\ref{2}) with respect to the de
Sitter vacuum solution $\bar R^{(0)}$ gives
\begin{eqnarray}
L'(R_0)\delta\hat R_{\mu\nu}+\frac{1}{4}[L''(R_0)R_0-2L'(R_0)]\bar
g^{(0)}_{\mu\nu}\delta\hat R-\frac{1}{2}L(R_0)h_{\mu\nu}
\\\nonumber +[\frac{R_0L'(R_0)}{8}-(\frac{R_0}{4})^2L''(R_0)]\bar
g^{(0)}_{\mu\nu}h=\bar\kappa^2 T_{\mu\nu}\label{14}
\end{eqnarray}

By Eq.(\ref{12}) and the fact that $L'(\hat R)$ is actually a
function of $T$ we can move all the "source" term in the above
equation to the right-hand side
\begin{eqnarray}
L'(R_0)\delta\bar R_{\mu\nu}+\frac{1}{4}[L''(R_0)R_0-2L'(R_0)]\bar
g^{(0)}_{\mu\nu}\delta\bar R-\frac{1}{2}L(R_0)h_{\mu\nu}
\\\nonumber +[\frac{R_0L'(R_0)}{8}-(\frac{R_0}{4})^2L''(R_0)]\bar
g^{(0)}_{\mu\nu}h=\bar\kappa^2
T_{\mu\nu}+L'(R_0)S_{\mu\nu}\label{15}
\end{eqnarray}
where $S_{\mu\nu}$ is a function of $\bar\nabla_\mu T$ and
$\bar\nabla_\mu\bar\nabla_\nu T$ given by
\begin{eqnarray}
S_{\mu\nu}=(L')^{-1}\bar\nabla_\mu\bar\nabla_\nu
L'+\frac{1}{2}(L')^{-1}\bar g^{(0)}_{\mu\nu}\bar\square
L'-\frac{3}{2}\bar\nabla_\mu L'\bar\nabla_\nu
L'\\\label{16}-\frac{3}{4}[\frac{L''(R_0)R_0}{L'(R_0)}-2]\bar
g^{(0)}_{\mu\nu}[\frac{1}{2}(L')^{-2}(\bar\nabla_\mu
L)^2-L'\bar\square L']\nonumber
\end{eqnarray}

It is easy to see that the choice of $L(R)=R-2\Lambda$ reproduces
the expansion equation of General Relativity around de Sitter
background \cite{Higuchi}, which is natural since now the Palatini
formulation is equivalent to the metric formulation. But note that
this is the most general Lagrangian that those two formulations
are equivalent.

Let us consider the gauge transformation
\begin{equation}
h_{\mu\nu}\longrightarrow
h_{\mu\nu}+\bar\nabla_\mu\xi_\nu+\bar\nabla_\nu\xi_\mu\label{17}
\end{equation}
where $\xi_\mu$ is an arbitrary vector field. By equation
$\bigtriangleup_{\xi}\delta \bar
R_{\mu\nu}=\frac{R_0}{4}(\bar\nabla_\mu\xi_\nu+\bar\nabla_\nu\xi_\mu)$,
it is not hard to check that Eq.(12) is invariant under this gauge
tansformation. Thus we can choose a suitable $\xi_\mu$ to impose
the transverse-traceless (TT) gauge on $h_{\mu\nu}$
\begin{equation}
\bar\nabla^\mu h_{\mu\nu}=0, h=0.\label{18}
\end{equation}
This gauge-invariance also implies that the graviton described by
theory (1) is still massless.

After gauge fixing, from Eq.(\ref{11}), the expansion equation
(\ref{16}) can be written in a much more simplified and
illuminating form
\begin{equation}
\bar\square
h_{\mu\nu}-\frac{R_0}{6}h_{\mu\nu}=\frac{-2\bar\kappa^2}{L'(R_0)}T_{\mu\nu}-2S_{\mu\nu}\label{19}
\end{equation}
If we identify the $\bar\kappa^2$ appearing in action (1) as
$\kappa^2/L'(R_0)$, where $\kappa^2=8\pi G$ and $G$ is the Newton
constant, this equation is identical to the expansion equation of
General Relativity around a de Sitter background except the
appearance of the $S_{\mu\nu}$ term. Note that for any reasonable
model, $L'(R_0)\sim 1$, thus $\bar\kappa$ is actually the same
order of magnitude as $\kappa$. Since $S_{\mu\nu}$ depends only on
the derivatives of $T$, for the vacuum or a constant $T$,
Eq.(\ref{19}) reduces exactly to the expansion equation of General
Relativity. This confirms in another way the conclusion that the
theory (1) will reduce to General Relativity in the case of vacuum
\cite{Volovich} and constant $T$ \cite{Rubilar} in an unified and
illuminating way.

Thus for a slowly varying source, denoting its density by
$\rho=-T$, the Newtonian potential it generates will be
\begin{equation}
\Phi(r)=\frac{G\rho-S_{00}/16\pi}{r}\exp(-\sqrt{\frac{R_0}{6}}r)\label{20}
\end{equation}
The exponential is just the standard Yukawa term in the
gravitational potential in de Sitter background in General
Relativity. With $R_0\sim H_0^2$, where $H_0$ is the current
Hubble parameter and is given by $H_0\simeq 100 km/s.Mpc$, the
effects of the Yukawa term can be neglected in solar system or
ground-based gravitational experiments. All the effects of the
modified action in Palatini formulation are enclosed in the
$S_{00}$ term. This is the secret of the modified gravity models
in Palatini formulation.

Since now the form of Newtonian potential has been severely
constrained by experiments (see Ref.\cite{Will} for a review), the
effects of $S_{00}$ should be extremely small to evade the current
constraints. Let us now estimate its magnitude for the class of
models that intend to explain the current cosmic acceleration. For
those models, we generally have $L(\hat R)=\hat R+f(\hat R)$. In
order to explain current acceleration and reduce to General
Relativity at early times, it requires that when $\hat R\sim
H_0^2$, the $f(\hat R)$ term dominates, when $\hat R\gg H_0^2$,
the $\hat R$ term dominates. Any test objects in current
gravitational experiments satisfy $\kappa^2\rho\gg H_0^2$, e.g.
for an object with density of the order $10^3 kg/m^3$,
$\kappa^2\rho/H_0^2\sim 10^{29}$. Thus in those cases, $L(\hat
R)\sim \hat R\sim \kappa^2\rho$ and $f(\hat R)\ll \hat R$. Then
\begin{eqnarray}
L'\sim 1+f(\hat R)/\hat R\sim 1,\ \ \bar\nabla_0L'=f''(\hat
R)\kappa^2\dot\rho\sim\frac{f(\hat R)}{\hat
R}\frac{\dot\rho}{\rho},\\\nonumber
\bar\nabla_0\bar\nabla_0L'=f''(\hat R)\kappa^2\ddot\rho+f'''(\hat
R)(\kappa^2\dot\rho)^2\sim \frac{f(\hat R)}{\hat
R}\frac{\ddot\rho}{\rho}+\frac{f(\hat R)}{\hat
R}(\frac{\dot\rho}{\rho})^2 \label{}
\end{eqnarray}
Thus we can see from Eq.(13) that all the terms in $S_{00}$ are
suppressed by a $f(\hat R)/\hat R$ factor, which is in practice
the order $H_0^4/(\kappa^2\rho)^2\sim 10^{-58}$ that is extremely
small for any test objects in current gravitational experiments.
Thus the Newtonian limit is always obtained. The above estimate is
quite obvious in a specific example: the $1/R$ gravity
\cite{CDTT}. There, $L(\hat R)=\hat R+\alpha^2/\hat R$, where
$\alpha\sim H_0^2$. $\hat R$ is related to $\rho$ trough
Eq.(\ref{scalar}) and is given by
\begin{equation}
\hat
R=\frac{1}{2}[\bar\kappa^2\rho+\sqrt{4\alpha^2+(\bar\kappa^2\rho)^2}]\label{21}
\end{equation}

The above computations and conclusions are also easily extended to
the matter loops corrected modified gravity in Palatini
formulation \cite{Wang4}. This type of models can be written as
\begin{equation}
S_{loop}[\bar{g}_{\mu\nu},
\hat{\bigtriangledown}_{\mu}]=\frac{1}{2\bar\kappa^2}\int
d^4x\sqrt{-\bar{g}}[\bar{R}+f(\hat{R})]\equiv
S_{EH}+S_{Palatini},\label{30}
\end{equation}
where $S_{EH}$ is the familiar Einstein-Hilbert action. When
written in the Einstein frame, the presence of a $\bar R$ term
will induce a kinetic term for the scalar field.

The counterpart of Eq.(11) is
\begin{eqnarray}
\delta\bar R_{\mu\nu}+f'(R_0)\delta\hat
R_{\mu\nu}+\frac{1}{4}[f''(R_0)R_0-2f'(R_0)]\bar
g^{(0)}_{\mu\nu}\delta\hat R-\frac{1}{2}[R_0+f(R_0)]h_{\mu\nu}
\\\nonumber +[\frac{R_0f'(R_0)}{8}-(\frac{R_0}{4})^2f''(R_0)]\bar
g^{(0)}_{\mu\nu}h=\bar\kappa^2 T_{\mu\nu}\label{}
\end{eqnarray}

Then following the same line of analysis above, we can see that
$\bar\kappa^2=\kappa^2/(1+f'(R_0))$ and the gravitational
potential is also given by Eq.(\ref{20}) with the $L'$ in the
expression for $S_{\mu\nu}$ replaced by $f'$. Thus in particular,
the model (\ref{30}) will reduce to General Relativity in the case
of vacuum or source with constant trace of energy-momentum tensor.

In conclusion, we computed the gravitational potential by weak
field expansion of Palatini formulation of modified gravity which
admits a de Sitter vacuum solution. We conclude that the Newtonian
limit is always obtained in those class of models and the
deviations from General Relativity is very small for slowly moving
source. It is well-known that there  are  some  motivations to
extend General Relativity, such as its incompatibility with
quantum mechanics, non-localities and unavoidable singularities in
its solutions. In the recent works the main motivation is to
explain the recently observed cosmic  acceleration without dark
energy. Although the proposal of a $1/R$ term is doomed now. It is
still worth continuing to explore whether some phenomenological
model of this type or its variants can easily accommodate the
data. There is an observation in this direction of model-building:
To drive an current cosmic acceleration, one only need the a de
Sitter vacuum solution and in order to reduce to General
Relativity, $L(R)\rightarrow R$ when $R/H_0^2\gg 1$. Specifically,
it is not necessary to have $L(R)\rightarrow\infty$ as
$R\rightarrow 0$ which may introduce instabilities.

\textbf{Acknowledgements}

We would especially like to thank Sergei Odintsov for a careful
reading of the manuscript and helpful comments. We also wish to
thank \'{E}.\'{E}.Flanagan and S.Nojiri for helpful discussions.
This work is partly supported by China NSF, Doctoral Foundation of
National Education Ministry and ICSC-World lab. scholarship.


\begin{thebibliography}{99}
\bibitem{Perlmutter} S. Perlmutter el al. Nature 404 (2000) 955;
Astroph. J. 517 (1999) 565; A. Riess et al. Astroph. J. 116 (1998)
1009; ibid. 560 (2001) 49; Y. Wang, Astroph. J. 536 (2000) 531;
D.N.Spergel, et al., astro-ph/0302207; L.Page et al.
astro-ph/0302220; M.Nolta, et al, astro-ph/0305097; C.Bennett, et
al, astro-ph/0302209;
\bibitem{CDTT} S.M.Carroll, V.Duvvuri, M.Trodden and M.
Turner, astro-ph/0306438; S. Capozziello, S. Carloni and A.
Troisi, "Recent Research Developments in Astronomy \&
Astrophysics" -RSP/AA/21-2003 [astro-ph/0303041]; S.Capozziello,
Int.J.Mod.Phys.D 11 (2002) 483;
\bibitem{Odintsov-string} S.Nojiri and S.D.Odintsov, hep-th/0307071;
\bibitem{Odintsov-R2} S.Nojiri and S.D.Odintsov, hep-th/0307288;
\bibitem{Odintsov-lnR} S.Nojiri and S.D.Odintsov, hep-th/0308176;
\bibitem{Vollick} D. N. Vollick, astro-ph/0306630;
\bibitem{Flanagan} \'{E}.\'{E}.Flanagan, astro-ph/0308111;
\bibitem{Wang1} X.H.Meng and P.Wang, to appear in Class. and Quant. Grav. [astro-ph/0307354]; ibid,
astro-ph/0308031;
\bibitem{Wang2} X.H.Meng and P.Wang, astro-ph/0308284;
\bibitem{Wang3} X.H.Meng and P.Wang, hep-th/0309062;
\bibitem{Flanagan2}  \'{E}.\'{E}.Flanagan, gr-qc/0309015;
\bibitem{Wang4} X.H.Meng and P.Wang, hep-th/0310038;
\bibitem{Chiba} T.Chiba, astro-ph/0307338;
\bibitem{Dick} R.Dick, gr-qc/0307052;
\bibitem{Soussa} M. E. Soussa and R. P. Woodard, astro-ph/0308114;
\bibitem{Dolgov} A. D. Dolgov and M. Kawasaki, astro-ph/0307285;
\bibitem{Odintsov-mini} S.Nojiri and S.D.Odintsov, hep-th/0310045;
\bibitem{Volovich} M.Ferraris, M.Francaviglia and I.Volovich,
Nouvo Cim. B108 (1993) 1313 [gr-qc/9303007]; ibid, Class.
Quant.Grav. 11 (1994) 1505;
\bibitem{Rubilar} G.F.Rubilar, Class. Quant. Grav. 15 (1998) 239;
\bibitem{Magnano} G.Magnano and L.M.Sokolowski, Phys.Rev.D {50}
(1994) 5039 [gr-qc/9312008];
\bibitem{Will} C. M. Will, Theory and Experiment in
Gravitational Physics, Cambridge Universtiy Press, Cambridge,
1993; Living Rev. Rel. 4 (2001) 4;
\bibitem{Higuchi} A. Higuchi, Nucl. Phys. B 325 (1989) 745;
\end{thebibliography}
\end{document}